\documentclass[12pt, draftclsnofoot, onecolumn]{IEEEtran}
\usepackage{balance}
\usepackage{cite, graphicx,amssymb,color,  pict2e, multirow, rotating} 
\usepackage{graphicx,amssymb}
\usepackage[tbtags]{amsmath}
\usepackage{times,amsmath}
\usepackage{subfig}
\usepackage{flushend}
\usepackage{amsmath}
\usepackage{epsfig}
\usepackage{amssymb}
\usepackage{hyperref}
\usepackage{color}
\usepackage{psfrag}
\usepackage{subfig}
\usepackage[usenames,dvipsnames]{xcolor}
\usepackage{textcomp}
\usepackage{array}
\usepackage{tikz}
\usetikzlibrary{matrix,decorations.pathreplacing}
\usepackage{booktabs}

\usepackage{tabularx,ragged2e}
\newcolumntype{C}{>{\Centering\arraybackslash}X} 

\usepackage{algorithmic}
\usepackage{algorithm}

\newtheorem{definition}{Definition}

\newtheorem{lemma}{Lemma}

\newtheorem{theorem}{Theorem}

\begin{document}
%
\title{MAC for Machine Type Communications in Industrial IoT -- Part I: Protocol Design and Analysis}
%
%
%

\author{
	    Jie Gao,~\IEEEmembership{Member,~IEEE,}
        Weihua Zhuang,~\IEEEmembership{Fellow,~IEEE,}
        Mushu Li,~\IEEEmembership{Student Member,~IEEE,}
        Xuemin (Sherman) Shen,~\IEEEmembership{Fellow,~IEEE,}
        and Xu Li
\thanks{
Jie Gao is with the Department of Electrical and Computer Engineering, Marquette University, Milwaukee, WI 53233 USA (e-mail: j.gao@marquette.edu).

Weihua Zhuang, Mushu Li, and Xuemin (Sherman) Shen are with the Department of Electrical and Computer Engineering, University of Waterloo, Waterloo, ON, Canada, N2L 3G1 (e-mail:
\{wzhuang, mushu.li\}@uwaterloo.ca, xshen@bbcr.uwaterloo.ca).

Xu Li is with Huawei Technologies Canada Inc., Ottawa,
ON, Canada, K2K 3J1 (email: Xu.LiCA@huawei.com).}
}

\maketitle

\begin{abstract}

In this two-part paper, we propose a novel medium access control (MAC) protocol for machine-type communications in the industrial internet of things. The considered use case features a limited geographical area and a massive number of devices with sporadic data traffic and different priority types. We target at supporting the devices while satisfying their quality of service (QoS) requirements with a single access point and a single channel, which necessitates a customized design that \textcolor{Black}{can significantly improve the MAC performance}. In Part~I of this paper, we present the MAC protocol that comprises a new slot structure, corresponding channel access procedure, and mechanisms for supporting high device density and providing differentiated QoS. A key idea behind this protocol is sensing-based distributed coordination for significantly improving channel utilization. To characterize the proposed protocol, we analyze its delay performance based on the packet arrival rates of devices. The analytical results provide insights and lay the groundwork for the fine-grained scheduling with QoS guarantee as presented in Part~II.

\end{abstract}


\IEEEpeerreviewmaketitle

\section{Introduction}

\IEEEPARstart{N}{ext}-generation wireless communications are envisioned to empower vertical industries~\cite{J_SShen_OJVT2020}. Among potential enterprise use cases, industrial communication networks for applications such as factory automation and process control are particularly important as they play a crucial role in Industrial 4.0, the upcoming fourth industrial revolution~\cite{J_SVitturi_Proc2019}. An indispensable step towards Industrial 4.0 is the development and standardization of industrial internet of things (IIoT), which connects devices such as sensors, actuators, and controllers to facilitate data collection and analysis and to support applications such as factory automation~\cite{S_GAceto_SurvTut2019}.

As in the general internet of things (IoT), machine-type communication (MTC), which facilitates automated data communications among devices, is a primary enabler of the IIoT. According to 3GPP, features of MTC include low mobility, small data packets, and time-controlled access~\cite{S_3GPP22368_2016}. Meanwhile, compared with the general IoT, IIoT has some unique characteristics: First, connectivity in IIoT is usually structured, featuring centralized network management and devices at fixed locations~\cite{J_ESisinni_TII2018}; Second, IIoT scenarios generally involve densely deployed devices in a relatively limited area. For example, the IIoT application of process monitoring may involve a density of up to $10,000$ devices per square kilometer~\cite{O_GBrown_WhitePaper2018}; Third, certain IIoT applications can be mission-critical and have extremely stringent quality of service (QoS) requirements. For example, the communication latency tolerance for machine tool motion control can be as low as 0.5 milliseconds (ms)~\cite{O_5GACIA_WhitePaper2018}.
The last two characteristics pose a significant challenge for supporting MTC in IIoT. Specifically, within a limited geographical area such as in a factory, the communication network may need to support a massive number of devices and, simultaneously, satisfy \textcolor{Black}{exceptionally strict} QoS requirements for some devices. 

In recent years, standards such as LTE-M, NB-IoT, and IEEE 802.11ah have been developed for supporting MTC. However, these standards usually focus on the general IoT instead of the IIoT use cases. As a result, they may not meet the stringent QoS requirements in IIoT, \textcolor{Black}{e.g.,  millisecond or sub-millisecond level delay}. For example, LTE-M and NB-IoT, both targeting at low power and long range communications, are more concerned with bandwidth usage and power consumption than latency. Specifically, the latency of LTE-M is no less than 10ms, while the latency of NB-IoT can be up to 10 seconds~\cite{S_SSharma_SurvTut2020}. The IEEE 802.11ah can support a less than 10ms latency but only under a low load condition, which limits its supported device density in practice~\cite{J_MAli_TVT2019}. 

To address challenges in supporting high device density and stringent QoS requirements, various solutions have been proposed for different layers in the network protocol stack. At the physical layer, exploiting the vast spectrum resource beyond 30GHz or even in the Terahertz band potentially provides support for a high device density~\cite{J_HSarieddeen_ArXiv2019}, \cite{M_XShen_Network2015}. In addition, physical-layer techniques, such as compressed sensing based multi-user detection, have been investigated for massive MTC (mMTC)~\cite{J_YDu_JSAC2017},\cite{J_GMA_TII2018}. However, the advances in the physical layer alone may not be sufficient for meeting the demand of MTC in IIoT for two reasons: First, the prevalence of IIoT relies on low-complexity and low-budget devices that may not support advanced physical-layer techniques; Second, even if physical-layer solutions can be applied, there is no guarantee that they will meet the stringent QoS requirements of IIoT. Therefore, developing a reliable link-layer solution becomes appealing as it is less limited by hardware and can be implemented with physical-layer solutions. 

There have been extensive research efforts in supporting MTC by link-layer design, despite few targets specifically at IIoT.  
For cellular-based MTC, the grant-based random access (RA) procedure for connection setup is a bottleneck that many MAC designs aim to address. When a large number of devices seek to set up connections in the RA procedure, the network can be congested~\cite{J_HMoussa_IoT2019}. MAC designs have been proposed with a focus on grouping and prioritizing devices, e.g., distributing devices into different groups after collisions~\cite{J_ABui_TVT2019}, grouping devices based on their delay requirements~\cite{J_RAbbas_TCom2017}, and prioritizing device transmissions using distributed binary sequences~\cite{J_M.Vilgelm_IoT2019}. \textcolor{Black}{3GPP release 15 includes the design of early data transmission (EDT), which replaces a standard four-step RA procedure with a two-step procedure~\cite{S_3GPP21915_2019}. While this change reduces delay~\cite{C_JThota_2019}, a grant-free access mechanism is still of great  interest~\cite{J_YLiu_JSAC_2020}.}~\footnote{\textcolor{Black}{3GPP release 16 also includes 5G support for IIoT through time sensitive communications (TSC)~\cite{S_3GPP21916_2020}. However, the focus of TSC is time synchronization instead of access control.}} After connection setup, delay in the data transmission phase can be reduced for all devices, via scalable transmission time intervals (TTI)~\cite{C_GPocovi_ICC2017},\cite{L_LYou_Comm2019} and, for high-priority (HP) devices in particular, via preempting scheduled low-priority (LP) transmissions~\cite{S_3GPPTR38.825_2019}.

For wireless local area network (WLAN) based MTC, many proposed MAC protocols focus on the improvement of 802.11ah. In 802.11ah, the mechanism for reducing transmission collision probability under high device density is the restricted access window (RAW), which divides devices into groups and uses time division in the channel access for different groups. Related MAC design efforts have been focused on the window size of RAW as well as the device grouping, e.g., optimizing the window size based on the number of devices~\cite{L_CPark_Comm2014}, adapting window size based on the number of transmission attempts in each group~\cite{J_YKim_TWC2016}, and using traffic-aware device grouping based on an estimation of channel usage in the groups~\cite{J_TChang_TMC2019}.

Beyond the preceding two categories of research works, which build on and improve existing solutions, more ambitious approaches have  been investigated. For example, machine learning based device-level traffic arrival forecast is adopted in~\cite{J_VRodoplu_IoT2020}, which simplifies MAC  into proactive channel scheduling. Another example is the re-configurable MAC proposed in~\cite{J_QYe_IoT2017} and \cite{J_AShoaei_TCOM2019} that dynamically adjusts the partition between contention-free and contention-based sections based on network traffic for maximizing network throughput.

The existing studies provides important insights on MAC  for MTC, such as the importance of coordinating and prioritizing devices and the necessity of a flexible design. However, for IIoT and in particular for applications such as factory automation and process control, further research on customized MAC protocols is necessary. In this paper, we focus on \textcolor{Black}{coordination-based grant-free} MAC as a link-layer solution to support such applications in IIoT. Specifically, we aim to achieve the following objectives: 1) to support a high device density in a limited geographical area, e.g., a manufacturing facility, with a single access point (AP) and a single channel; 2) to provide differentiated QoS to different types of devices, while satisfying  stringent QoS requirements of HP devices, \textcolor{Black}{e.g., millisecond or sub-millisecond level delay}; 3) to minimize messaging and control overhead; 4) to accommodate devices with simple and low-cost hardware, i.e., without requiring advanced physical-layer techniques. Achieving all the objectives simultaneously requires a protocol that fully exploits the potential of MAC design. We present our proposed MAC solution in two parts, with details of a new MAC protocol in Part~I and a customized scheduling scheme to complement the protocol in Part~II~\cite{J_JGao_JIoT_2020PII}. Through the two parts, the integration of delicate distributed coordination and fine-grained centralized scheduling composes the unique strength of our MAC solution.

The contribution of Part~I is two-fold:
First, we propose a novel MAC protocol for applications in IIoT such as factory automation and process control. The protocol design comprises a new time slot structure,  corresponding channel access procedure, and two mechanisms for providing differentiated QoS and for supporting ultra-dense networks, respectively. Featuring delicate distributed coordination, the protocol can significantly improve channel usage efficiency without packet collision or, in the case of high device density, support a large number of devices at the cost of low packet collision probabilities; 
Second, we provide thorough performance analysis for the proposed MAC protocol based on limited data traffic information. Specifically, we characterize the delay performance for the proposed channel access strategy as well as the impact of the two mechanisms. Without assuming a specific traffic arrival model, we establish our analysis  based only on the packet arrival rates at the devices. The analytical results provide insight for scheduling and are later demonstrated to be accurate by simulations in Part~II.

The rest of Part~I is organized as follows. Section~\ref{s:Scenario} describes the networking scenario under consideration. Section \ref{s:ProtoclDesign} presents the proposed MAC protocol. In Section~\ref{s:PeformAnalysis}, we provide performance analysis for the proposed MAC design. Section~\ref{s:Conclude} concludes Part~I. Proofs of the theorems are given in Appendix. A list of main symbols is given in Table~\ref{t:Notation}.


\section{Networking Scenario}\label{s:Scenario}

\begin{figure}[tt!]
	\vspace{-2mm}
	\begin{centering}
		\includegraphics[width=0.5\textwidth]{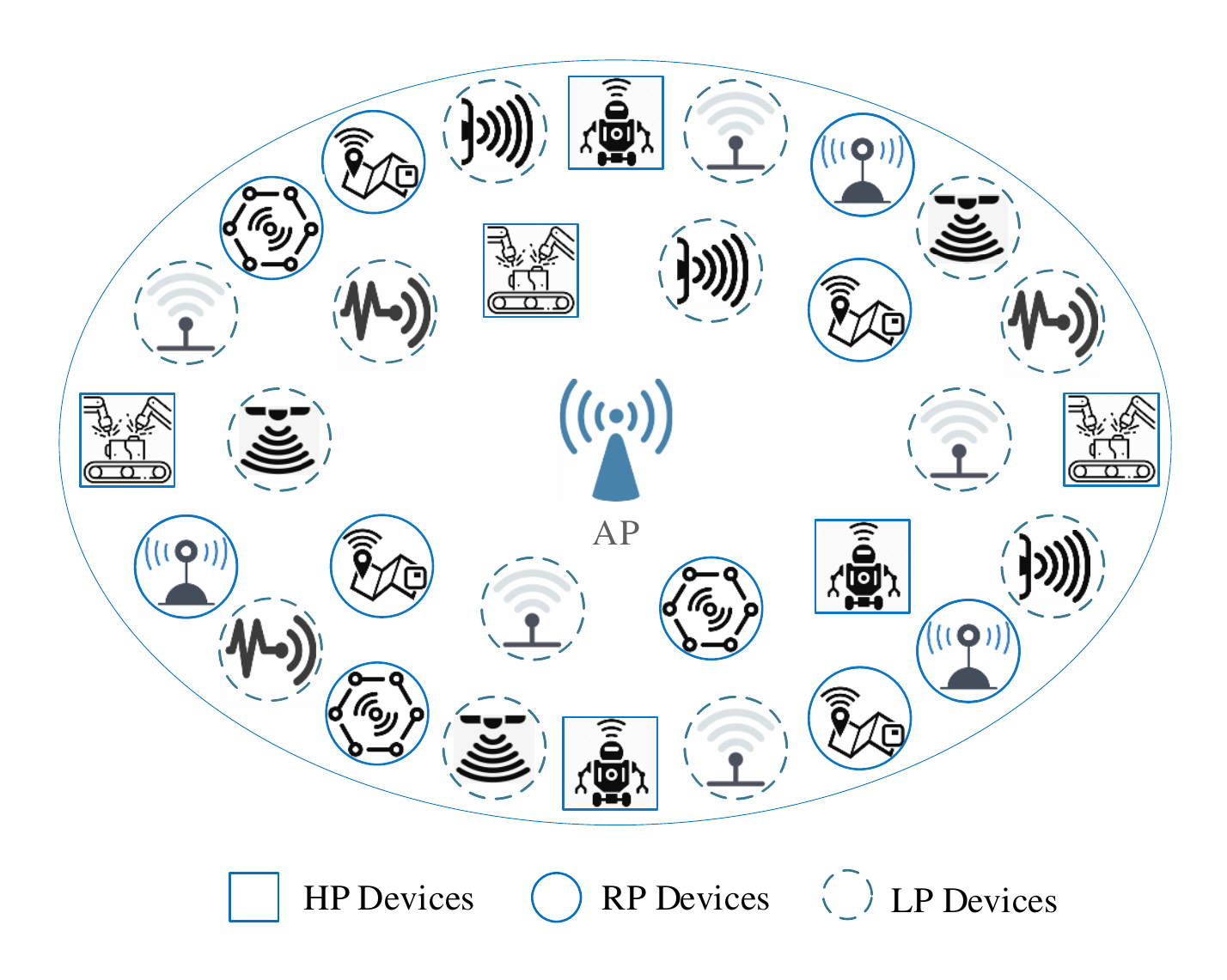}
		\par\end{centering}
	\vspace{-2mm}
	\centering{\caption{An illustration of the networking scenario.}\label{f:Sys}}
\end{figure}

Consider a fully connected network with one AP covering a limited geographical area, e.g., a manufacturing facility.~\footnote{The target area is assumed to be less than 1 km$^2$.} A large number of devices such as sensors, actors, and controllers are densely deployed in the area. The devices are categorized into three types, i.e., HP devices, regular-priority (RP) devices, and LP devices. An illustration of the considered scenario is given in Fig.~\ref{f:Sys}.

The overall number of devices and the set of devices are denoted by $D$ and $\mathcal{D}$, respectively. The number and set of HP, RP, and LP devices are denoted by $D^\mathrm{H}$ and $\mathcal{D}^\mathrm{H}$, $D^\mathrm{R}$ and $\mathcal{D}^\mathrm{R}$, and $D^\mathrm{L}$ and $\mathcal{D}^\mathrm{L}$, respectively. Without loss of generality, we assume that the devices are indexed such that the first till the $D^\mathrm{H}$th devices are the HP devices, the next $D^\mathrm{R}$ devices are the RP devices, and the last $D^\mathrm{L}$ devices are the LP devices.

\textit{Communication Characteristics}. The communication characteristics include:
\begin{itemize}
	\item Short data packets - The length of physical-layer packets is normally in the range between several bytes to several hundred bytes~\cite{M_DKim_Net2017};
	\item Uplink-dominated transmission - A significant portion of the data traffic is attributed to sensor readings or device status reports~\cite{M_CBockelmann_Com2016}.
\end{itemize}


\textit{QoS Requirements}. The considered QoS metrics are delay,  from the instant of packet arrival to the instant of successful packet transmission, and packet transmission collision probability. Different types of devices have different QoS requirements. Specifically, the maximum tolerable delay and packet collision probability for HP, RP, and LP devices are denoted by $\delta^\mathrm{H}$ and $\rho^\mathrm{H}$, $\delta^\mathrm{R}$ and $\rho^\mathrm{R}$, and $\delta^\mathrm{L}$ and $\rho^\mathrm{L}$ respectively, where $\delta^\mathrm{H} < \delta^\mathrm{R} < \delta^\mathrm{L}$ and $\rho^\mathrm{H} < \rho^\mathrm{R} < \rho^\mathrm{L}$. The value of $\delta^\mathrm{H}$ is assumed to be  small such as on the millisecond level.

\textit{Device Packet Arrivals}. For practicality, we do not assume a specific traffic model. However, we consider the following data packet arrival properties:
\begin{itemize}
	\item The packet arrival statistics at each device are constant during a relatively long period with respect to packet inter-arrival time. The packet arrival rate of device $i$ in the considered time duration is denoted by $\lambda_{i}$;
	\item The packet arrival rate is relatively low so that $1/\lambda_{i}$ is much larger than $\delta^\mathrm{H}$ for any $i$. This is in accordance with the sporadic transmission characteristic of machine type communications, where the packet inter-arrival time can range from tens of milliseconds to several minutes~\cite{S_SSharma_SurvTut2020}; 
	\item For tractability, we assume that the transmission time for data packets is identical and equal to $T_\mathrm{x}$.
\end{itemize}



Given the networking scenario, we aim to develop a MAC solution with the following features:
\begin{enumerate}
	\item Accommodating a large number of devices on a single channel with a single AP;
	\item Satisfying the differentiated QoS requirements for each type of devices;
	\item Keeping control overhead as low as possible;
	\item Exploring the role of machine learning, specifically in device transmission scheduling.
\end{enumerate}
In Part~I, we focus on items 1), 2), and 3), while in Part~II our emphasis is placed on items 1), 2), and 4).



\begin{table}[t!]
	\begin{center}
		\caption{\textcolor{Black}{List of Main Symbols}}\label{t:Notation}
		{\setlength{\extrarowheight}{1.55pt}
			\begin{tabularx}{0.9\textwidth}{c|X}\hline\hline
				$d_{m,l}$ & the index of the device assigned mini-slot $m$ of slot $l$ \\ \hline
				$D$ & the number of all devices \\ \hline
				$\mathcal{D}$   &  the set of all devices \\ \hline
				$\!\!\!\mathcal{D}^\mathrm{H}/\mathcal{D}^\mathrm{R}/\mathcal{D}^\mathrm{L}\!\!\!\!$ &  the set of all HP/RP/LP devices \\ \hline
				$\mathcal{D}_{m,l}$ & the set of devices assigned mini-slot $m$ of slot $l$ \\ \hline
				$\mathcal{D}_l$ & the set of devices assigned slot $l$ \\ \hline
				$n_\mathrm{m}$ & the number of mini-slots in a slot \\ \hline
				$n_\mathrm{s}$ & the number of slots in a frame \\ \hline
				
				$r^\mathrm{H}/r^\mathrm{R}/r^\mathrm{L}$ & the number of slots in a HP/RP/LP assignment cycle \\ \hline
				$T_\mathrm{m}$ & the length of a mini-slot  \\ \hline
				$T_\mathrm{s}$ & the length of a slot  \\ \hline
				$T_\mathrm{x}$ & the length of a packet transmission duration \\ \hline
				
				$\delta^\mathrm{H}/\delta^\mathrm{R}/\delta^\mathrm{L}$ & the maximum tolerable delay of HP/RP/LP devices \\ \hline
				$\lambda_i$ & the packet arrival rate of device $i$  \\ \hline
				$\lambda_{m,l}$ & the packet arrival rate of device $d_{m, l}$ \\ \hline
				$\lambda_{m,l}^\prime$ & the effective packet arrival rate of device $d_{m, l}$ \\ \hline
				$\rho^\mathrm{H}/\rho^\mathrm{R}/\rho^\mathrm{L}$ &  the maximum tolerable packet collision probability of HP/RP/LP devices \\ \hline
				$\tau_{m,l}$ & the AD-F for the device assigned mini-slot $m$ of slot $l$ \\ \hline
				$\tau_{m,l}^\mathrm{b}$ & the AD-F for the device assigned mini-slot $m$ of slot $l$ in the case with buffer \\ 
				\hline\hline
			\end{tabularx}
		}
	\end{center}
	\vspace{-0.2cm}
\end{table}

\section{Proposed MAC Protocol}\label{s:ProtoclDesign}

The proposed MAC protocol is based on time-slotted channel access, which suits short packets. Tailored for the considered networking scenario, our
protocol comprises the following  elements:
\begin{itemize}
	\item Mini-slot based carrier sensing (MsCS);
	\item Synchronization carrier sensing (SyncCS);
	\item Differentiated assignment cycles;
	\item \textcolor{Black}{Superimposed} mini-slot assignment (SMsA).
\end{itemize}
In the list, the first two elements target at improving channel utilization efficiency through implicit distributed coordination, the third targets at providing differentiated QoS for different device types, and the last targets at increasing the number of supported devices.


\subsection{Time Frame and Slot Structure}

\begin{figure}
	\begin{centering}
		\vspace{-2mm}
		\includegraphics[width= 0.42\textwidth]{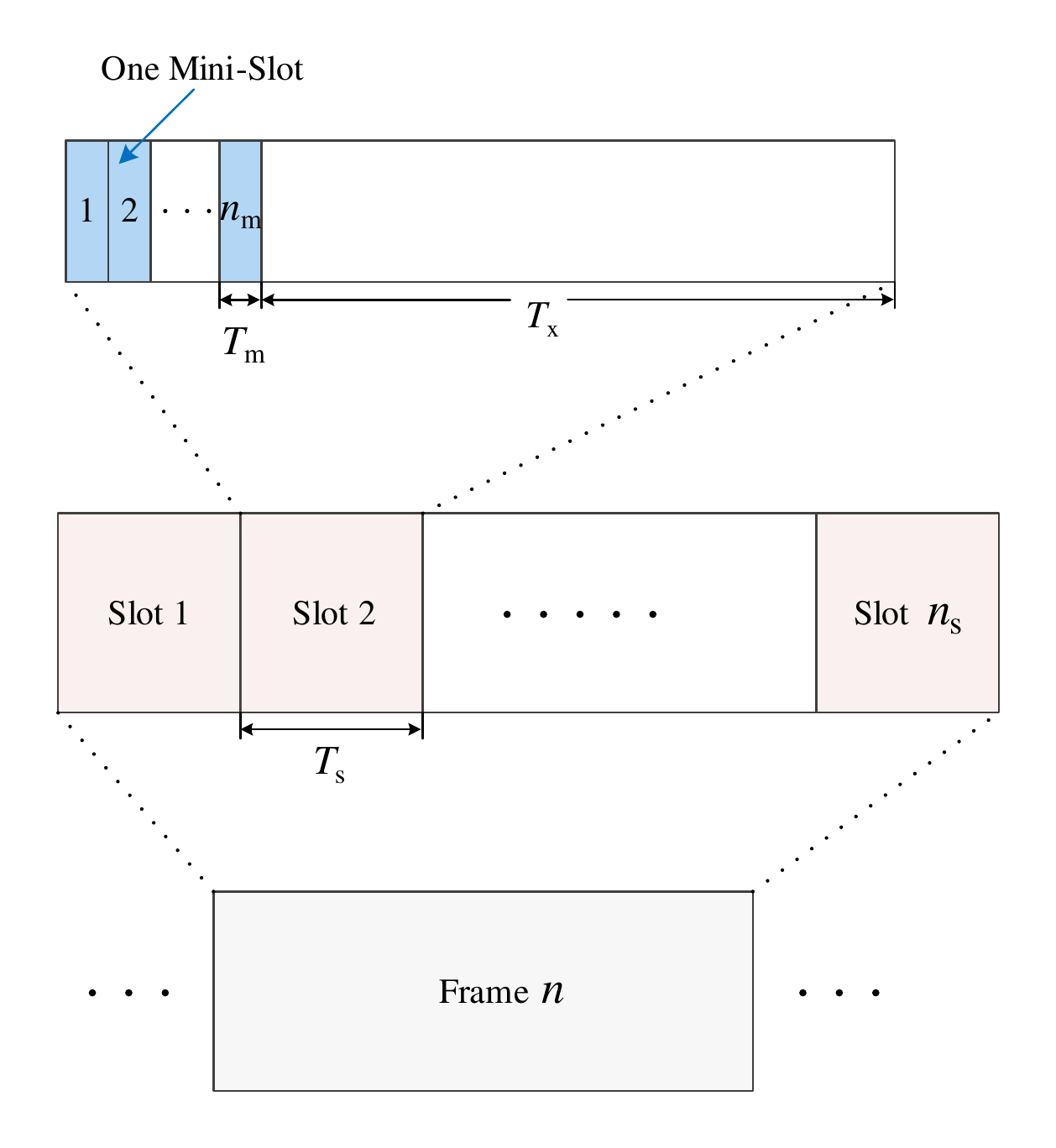}
		\par\end{centering}
	\centering{\caption{An illustration of the frame, slot, and mini-slot  structure.}\label{f:Slot}}
\end{figure}

Time is partitioned into frames, and each frame is partitioned into $n_\mathrm{s}$ slots,  as shown in Fig.~\ref{f:Slot}.  A slot begins with $n_\mathrm{m}$ mini-slots, each of length $T_\mathrm{m}$, followed by a duration of length $T_\mathrm{x}$. Accordingly, the length of a slot, denoted by $T_\mathrm{s}$, depends on the number of mini-slots and is equal to $n_\mathrm{m}\times T_\mathrm{m} + T_\mathrm{x}$. 

 Given the high device density and sporadic transmission pattern, each slot is assigned to multiple devices, in order to achieve high channel utilization efficiency via reducing idle slots. Different devices associated with a slot are assigned different mini-slots of the slot. Different from existing designs with mini-slots (e.g., \cite{J_LAlonso_JSAC2000}, \cite{J_PWang_TWC2009}, or \cite{J_HDhillon_TCOM2014}), where each mini-slot is used for transmitting one or more packets, the mini-slots in our protocol are very short (e.g., less than 10 $\mu$s) and are used for channel sensing \textcolor{Black}{instead of sending reservation requests or data packets} (as detailed in Subsection~\ref{ss:MsCs}). In the proposed protocol, the minimum time unit for transmitting a packet is a slot, and each slot accommodates at most one successful packet transmission. Clearly, without proper coordination, transmission collision may happen when multiple devices are assigned to the same slot.

\subsection{MsCS}\label{ss:MsCs}

\begin{figure}
	\centering
	\subfloat[Devices assigned to the first mini-slot of any slot  starts transmission immediately when the slot begins, without sensing the channel.]
	{\includegraphics[width=0.48\textwidth]{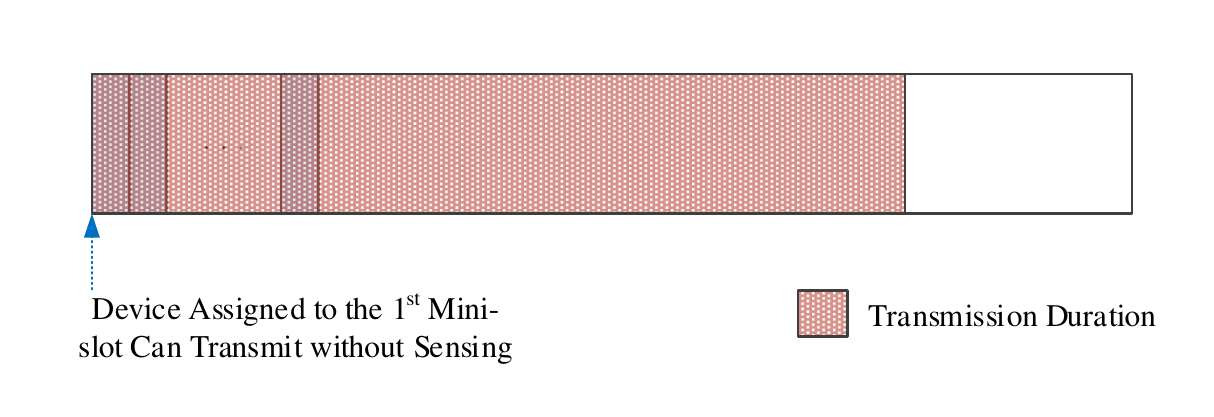}} 
	\hspace{1mm} 
	\subfloat[Devices assigned to mini-slot $m (>1)$ must sense the channel during the $(m-1)$th mini-slot, and starts transmission at the beginning of the $m$th mini-slot if the channel is sensed to be ilde.]
	{\includegraphics[width=0.48\textwidth]{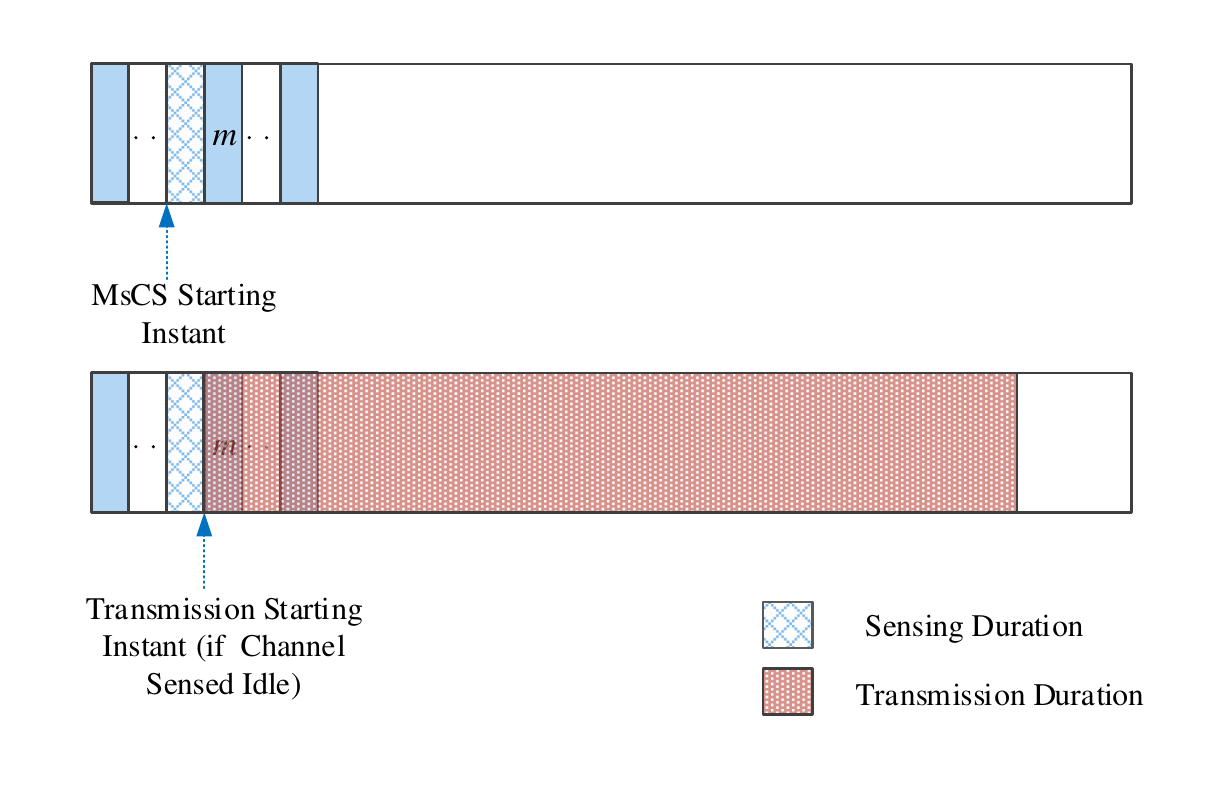}} 
	\caption{An illustration of the MsCS. }\label{f:Minislot}
\end{figure}

The purpose of mini-slots is to enable channel sensing for collision-free distributed channel access. When the AP assigns a slot to a device, it also specifies a mini-slot for the device. Suppose that device $i$ is assigned mini-slot $m$ of slot $l$ (such assignment is abbreviated as $\{l, m\}_i$ in the sequel). Then, the following rules are used in the proposed protocol:
\begin{itemize}
	\item If device $i$ has a packet to transmit and $m = 1$, it starts transmitting right away when slot $l$ begins;
	\item If device $i$ has a packet to transmit and $m > 1$, it needs to sense the channel during mini-slot $m-1$ of slot $l$ and starts transmitting from mini-slot $m$ of slot $l$ if the channel is sensed to be idle;
	\item If device $i$ does not have a packet to send, it simply stays idle in the corresponding slot.
\end{itemize}
The first two cases are illustrated in Fig.~\ref{f:Minislot}.

With MsCS, different mini-slots correspond to different transmission priorities. Specifically, a mini-slot with a larger index corresponds to a lower transmission priority. Therefore, mini-slots with small indexes can be used to accommodate HP devices. Through MsCS, before accessing the channel, a device makes sure that none of the devices with higher priority is using the channel. As a result, the devices can avoid packet collision while sharing the same slot. Note that the MsCS is fully distributed and does not require any control message exchange, given the assignment of slots and mini-slots to devices by the AP. The cost for avoiding collision is the overhead of using mini-slots for sensing. Specifically, the ratio of packet transmission duration over slot length is $T_\mathrm{x}/T_\mathrm{s}$.


For MsCS to work, the following conditions should be satisfied:
\begin{itemize}
	\item The mini-slot length, $T_\mathrm{m}$, must be longer than the maximum propagation delay across the network coverage area;~\footnote{\textcolor{Black}{A possible choice for mini-slot length is 9$\mu$s, which follows from the distributed coordination function (DCF) slot time in IEEE~802.11ac.}}
	\item The overall length of all mini-slots, i.e.,  $n_\mathrm{m}T_\mathrm{m}$, should be less than the packet transmission duration $ T_\mathrm{x}$ (for each slot to accommodate at most one transmission);\footnote{In an extreme case when an HP device transmits a very short packet, it is possible that an LP device senses channel idle and transmits a packet in the same slot. This extreme case is ignored in the protocol performance analysis.}
	\item The aggregated packet arrival rate of all devices assigned the same slot must be less than 1 per frame.
\end{itemize}

%
%
%
%

\subsection{SyncCS}

Even though MsCS improves channel usage efficiency, as a result of multiple devices sharing each slot, none of the devices may have a packet to transmit in a slot. Increasing the number of mini-slots in each slot can reduce the slot idle probability. However, it may violate the delay requirements for devices assigned high-index mini-slots or the aforementioned condition on the overall length of mini-slots.

\begin{figure}
	\vspace{-4mm}
	\centering
	\vspace{-1mm}
	\subfloat[When the last mini-slot of a slot is sensed idle, the remaining transmission duration of this slot is skipped, and the next slot starts right after the last mini-slot of this slot.]
	{\includegraphics[width=0.49\textwidth]{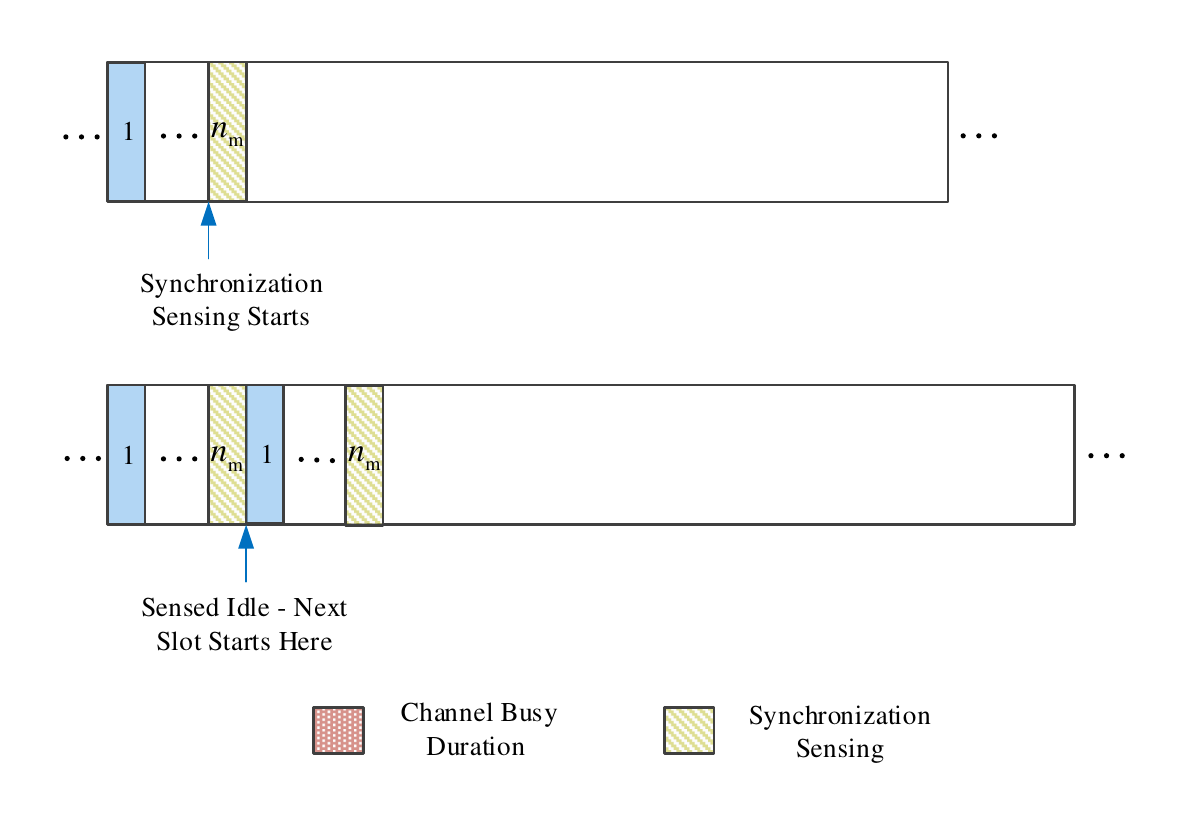}\label{f:SyncSense_a}} 
	\vspace{-1mm}
	\subfloat[When the last mini-slot of a slot is sensed busy, the next slot starts after the entire duartion of this slot.] 
	{\includegraphics[width=0.49\textwidth]{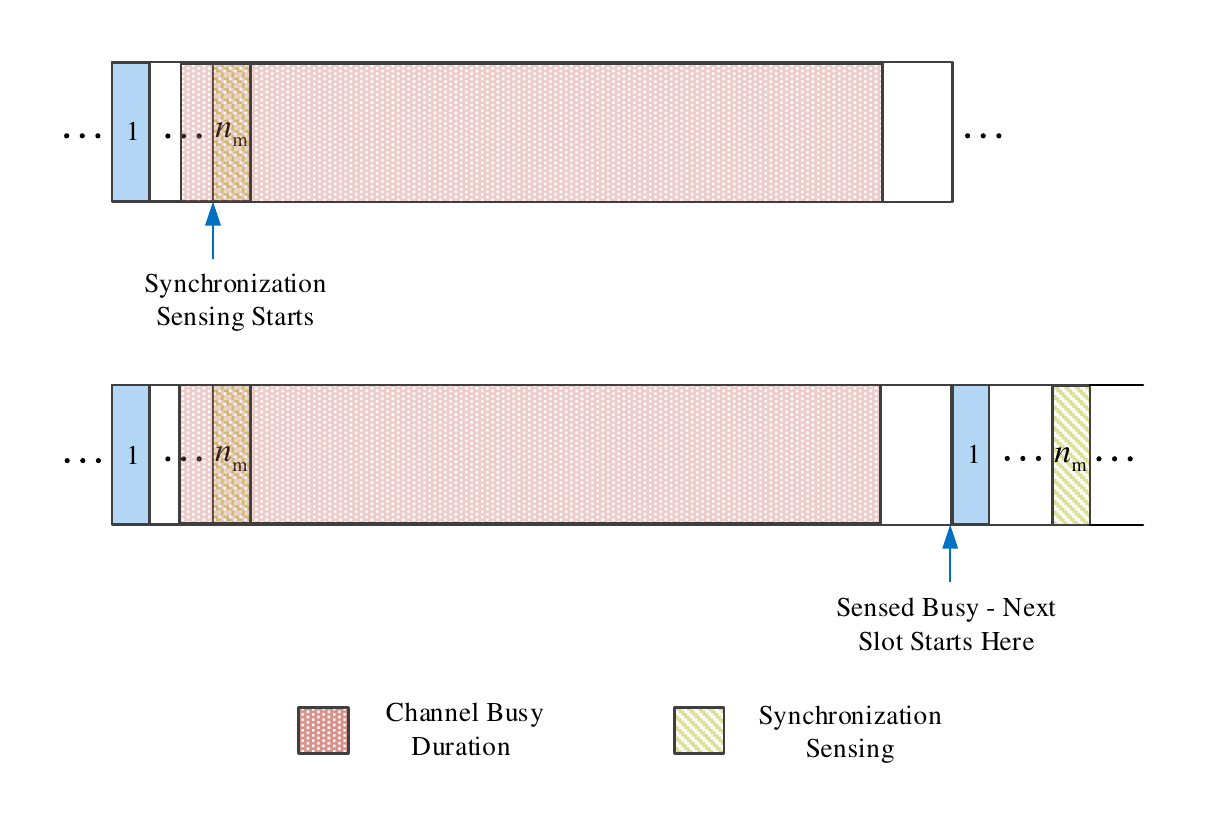}\label{f:SyncSense_b}} 
	\caption{An illustration of the syncCS.}\label{f:SyncSense}
\end{figure}

Alternatively, if idle slots can be identified and avoided, the channel usage efficiency can be further improved, and so will the resulting QoS. To achieve this, the following rules of SyncCS are used in the proposed protocol:
\begin{itemize}
	\item All devices in $\mathcal{D}$ sense the channel in the last mini-slot, i.e., mini-slot $n_\mathrm{m}$, of everyone slot. The only exception is the device that is assigned mini-slot $n_\mathrm{m}$ and has a packet to transmit from that mini-slot;~\footnote{The device that has a packet to transmit and is assigned mini-slot $n_\mathrm{m}$ of a slot starts to transmit from mini-slot $n_\mathrm{m}$ after sensing the channel idle in all the previous $n_\mathrm{m}-1$ mini-slots.}
	\item If the last mini-slot is idle, the rest of the current slot is skipped and the next slot starts immediately after this last mini-slot;
	\item If the last mini-slot is busy, the next slot starts after the current slot ends.
\end{itemize}
The above rules are illustrated in Fig.~\ref{f:SyncSense}, and the rationale is explained as follows. Given the condition that $n_\mathrm{m}T_\mathrm{m} < T_\mathrm{x}$ as mentioned in Subsection ~\ref{ss:MsCs}, no device is or will be transmitting in a slot if the last mini-slot of that slot is idle. Therefore, upon sensing an idle last mini-slot, all devices know that the rest of the slot can be skipped and
the next slot can start after this mini-slot. The SyncCS allows devices to synchronize slots even though the length of a slot is no longer fixed. With SyncCS, a busy slot has the full length of  $n_\mathrm{m}\times T_\mathrm{m} + T_\mathrm{x}$, while an idle slot has the reduced length of $n_\mathrm{m}\times T_\mathrm{m}$. 

The SyncCS has two main differences from the MsCS:
\begin{itemize}
	\item In SyncCS, devices must perform sensing regardless of whether they have a packet to transmit or not (with the only exception as mentioned above);
	\item In SyncCS, all devices, not just the devices assigned to the slot, need to sense the channel in each slot.
\end{itemize}


Similar to the MsCS, SyncCS is fully distributed and does not require any control message exchange. The cost for further improving channel utilization efficiency via SyncSC is the extra channel sensing. In addition, accurate time synchronization is required among all devices. Without SyncSC, a device can be in the sleep mode for most of the time in a frame and only wake up before its assigned mini-slot for MsCS if it has a packet to transmit. With SyncSC, each device needs to perform sensing in each slot \textcolor{Black}{and re-synchronize once for each idle slot}. In the IIoT scenario under consideration, it is possible that energy consumption of devices is less of a concern (e.g., as compared with sensors deployed in remote areas such as in forests); Otherwise, the design element of SyncSC can be omitted in the proposed protocol.~\footnote{\textcolor{Black}{Alternatively, the AP may broadcast frame synchronization beacons. In such case, when a device has a packet to send, it can wake up and synchronize to the next frame. It may remain awake and synchronized to each slot until the packet is transmitted.}}


\subsection{Differentiated Assignment Cycles}\label{ss:DiffAssign}

\begin{figure}
	\vspace{-4mm}
	\begin{centering}
		\includegraphics[width=0.5\textwidth]{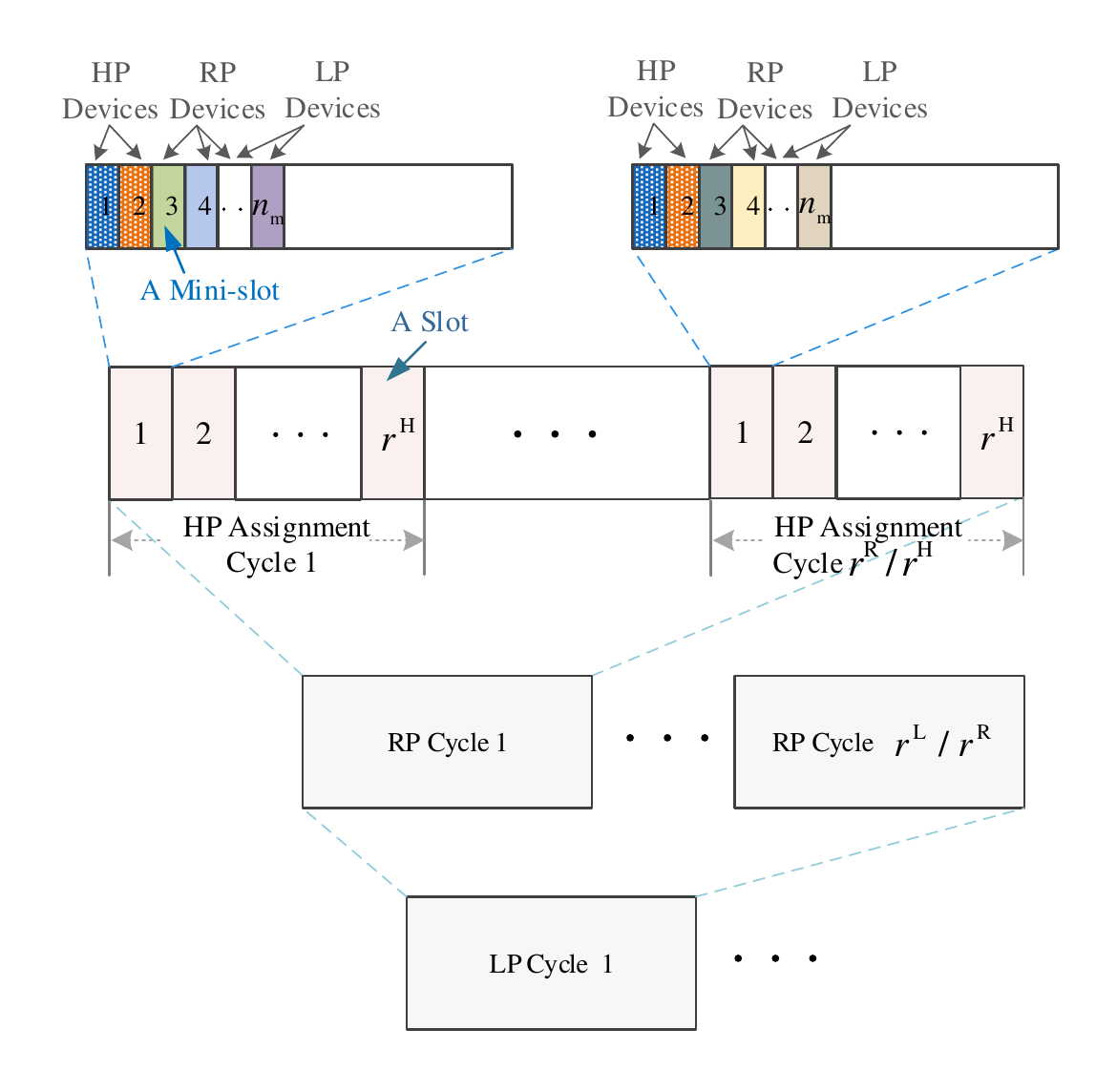} 
		\par\end{centering}
	\centering{\caption{An illustration of differentiated assignment cycles.}\label{f:HPCycle}}
\end{figure}

Using the slot structure in Fig.~\ref{f:Slot},  the delay for a device depends on the frame length if each device has at most one transmission opportunity in each frame. However, one transmission opportunity in each frame for every device does not provide sufficient flexibility to support differentiated QoS. Particularly, the maximum delay threshold of HP devices, i.e., $\delta^\mathrm{H}$, can be much smaller than that of RP/LP devices.
To address this problem, we extend the frame in Fig.~\ref{f:Slot} to differentiated assignment cycles. Specifically, each HP, RP, and LP assignment cycle consists of $r^\mathrm{H}$, $r^\mathrm{R}$, and $r^\mathrm{L}$ slots, respectively, where $r^\mathrm{H} < r^\mathrm{R} < r^\mathrm{L}$. Each HP, RP, or LP device is assigned one mini-slot of one slot in each HP, RP, or LP assignment cycle, respectively. \textcolor{Black}{Thus, an HP/RP/LP cycle serves as a frame for the HP/RP/LP devices, respectively. In the case when all devices have the same priority, the HP, RP, and LP cycles become identical and reduce to a standard frame.}
The differentiated assignment cycles are illustrated in Fig.~\ref{f:HPCycle}, in which different color patterns in the mini-slots represent different assigned devices. In the illustration, $r^\mathrm{L}$ is a multiple of $r^\mathrm{R}$, and $r^\mathrm{R}$ is a multiple of $r^\mathrm{H}$.~\footnote{While $r^\mathrm{L}$ does not have to be a multiple of $r^\mathrm{R}$ or $r^\mathrm{H}$ in theory, the overall device assignment cycle is the lowest common multiple of $r^\mathrm{H}$, $r^\mathrm{R}$, and $r^\mathrm{L}$. Limiting the lowest common multiple to be $r^\mathrm{R}$ itself can reduce the complexity of device assignment by the AP.} The HP devices assigned to the same slot in any different HP assignment cycles are identical, as shown by the two illustrated slots on the top of Fig.~\ref{f:HPCycle}, while the RP or LP devices assigned to the two slots are different. 

With differentiated assignment cycles, it becomes possible to achieve the stringent delay requirement of HP devices, by setting $r^\mathrm{H}$ small, and at the same time support a large number of devices, by using a large $r^\mathrm{R}$ and/or $r^\mathrm{L}$. 
Note that similar idea of differentiated cycles can be found in existing work such as \cite{J_KMalekshan_TWC2014}, where two different cycle lengths are used for realtime and non-realtime traffic, respectively. With a different slot structure and three different cycle lengths, we adopt the same essential idea here. This is because, for scheduling based channel access, achieving lower delay translates to more frequently scheduled transmission opportunities. This naturally leads to differentiated cycles for different device or traffic types.


\subsection{SMsA}

The proposed MAC protocol aims to support a high device density. The MsCS and SyncCS contribute to the solution by improving channel utilization efficiency, along with differentiated assignment cycles   with a large $r^\mathrm{R}$ and/or $r^\mathrm{L}$. In addition, if devices can share a mini-slot, beyond only sharing a slot, the capacity of the network in terms of the number of supported devices can be significantly improved, at the cost of nonzero packet transmission collision probabilities.

The final element in our proposed protocol, i.e., SMsA, allows the assignment of one mini-slot to multiple devices, provided that packet transmissions associated with such assignment can be properly scheduled as not to
violate the QoS requirements of the devices. For simplicity in presentation, we limit the SMsA to devices of the same type, i.e., an HP device can share a mini-slot only with other HP devices. With SMsA, a mini-slot in Fig.~\ref{f:HPCycle} may no longer be assigned to a device exclusively.

Transmission collision may happen among devices sharing a mini-slot, and the collision probability depends on the following factors:
\begin{itemize}
	\item The device packet arrival rates;
	\item The number of mini-slots and the mini-slot assignment;
	\item The HP, RP, LP assignment cycle lengths.
\end{itemize}
While the device packet arrival rates are not controllable, the collision probability may be reduced by properly determining the last two factors (as presented in Part~II).

\textcolor{Black}{We do not consider collision resolution in this work. However, a design element for collision detection can be added in our proposed MAC protocol. The following is an example. If two or more devices assigned the same mini-slot simultaneously start sending packets to the AP, the AP will detect the collision. As soon as the AP detects the collision, it will start broadcasting a collision beacon that fills the rest of the current slot. On the device side, the sending devices will switch to sensing mode to check for a collision beacon after transmitting their packets. If a beacon is sensed, the device knows that a packet collision happened during its transmission and may decide to re-transmit the packet.}


%
%

%

\subsection{Downlink Control}

The AP broadcasts the mini-slot and slot assignment to devices via downlink control messages.  Based on the assumption of stationary traffic statistics in a relative long duration~\footnote{The stationary duration, if denoted by $T_\mathrm{st}$, should satisfy  $T_\mathrm{st} \gg r^\mathrm{L} T_\mathrm{s}$.}, the assignment does not need to be updated frequently. The AP may either broadcast the entire assignment in one downlink control message or breakdown the assignment information into multiple messages.

Consider an example of 10 mini-slots per slot (i.e., $n_\mathrm{m} = 10$) and 200 slots per LP assignment cycle (i.e., $r^\mathrm{L}$ = 200). In such case, 2 bytes is more than sufficient to represent the slot and mini-slot assignment for each device. For 1000 devices, the assignment message payload size is no more than 2 kilobyte (KB). For a slot length of 200 $\mu$s, an LP assignment cycle is about 40ms in length. Even if the traffic statistics change as frequently as once in every five minutes, the 2KB downlink assignment message is needed just once in every 7500 LP assignment cycles or, equivalently, $1.5\times 10^{6}$ slots.

As downlink control messages are infrequent in comparison with the dominating uplink messages, we neglect the impact of downlink control messages while analyzing performance of the proposed protocol.\\

To summarize, the core of MAC protocol is how to coordinate transmissions from devices, while prioritizing and device grouping are two important aspects of coordination. There are various approaches for prioritizing, such as using both contention-based and contention-free access in a MAC protocol~\cite{J_KMalekshan_TWC2016}. Similarly, there are many grouping approaches, such as limiting contention to devices generating packets around the same instant~\cite{J_JGao_TVT2018}. In our proposed MAC protocol, the utilization of mini-slots is inherently capable of both prioritizing and grouping. Meanwhile, the differentiated assignment cycles further strengthen the design's capability in prioritizing, while the SMsA further strengthens the capability in grouping. 

\section{Performance Analysis}\label{s:PeformAnalysis}

In this section, we present performance analysis of the proposed MAC protocol, focusing on the MsSC, SyncSC, and SMsA. Note that the proposed MAC protocol works under the following conditions:
\begin{itemize}
	\item The expected number of packet arrivals for all devices sharing a slot is less than 1 per frame.~\footnote{In the case with differentiated assignment cycles, the limit of `less than 1 per frame' should be replaced by `less than 1 per LP assignment cycle', and the $T_\mathrm{f}$ in \eqref{e:GenCond1} should be replaced with $r^\mathrm{L}T_\mathrm{s}$.}
	\item The average packet arrival interval of any device is larger than \textcolor{Black}{the maximum tolerable packet delay} of that device.
\end{itemize}
In practice, some devices can have a high packet arrival rate that violates the above conditions. In such case, more than one slot can be assigned to such a device in the corresponding assignment cycle so that the expected number of packet arrivals of the device per scheduled slot is less than one. In the subsequent analysis, we simply assume that the number of packet arrivals for any device is less than one per its assignment cycle.


Without assuming a specific traffic model, we focus on the first-order statistic. The expected number of packet arrivals at device $i$ in a frame is given by $\lambda_{i}T_\mathrm{f}$, where $T_\mathrm{f}$ denotes the length of a frame. Denote the set of all HP, RP, and LP devices assigned to slot $l$ by $\mathcal{D}_l$. Denote the delay of device $i$, averaged over packet transmissions while the traffic is stationary, by $\tau_i$. The aforementioned two conditions correspond to the following equations:
\begin{subequations}
\begin{align}
\sum\limits_{i \in \mathcal{D}_l} \lambda_{i}  T_\mathrm{f}  \leq 1, \forall l,  \label{e:GenCond1}\\
\tau_i  \leq \frac{1}{\lambda_{i}}, \forall i. \label{e:GenCond2}
\end{align}
\end{subequations}

\subsection{Delay Performance with No Buffer}\label{ss:minislotDelay}

We investigate the impact of mini-slots in the case without SMsA, given the slot assignment and device packet transmission probabilities (estimated from the packet arrival rates). Starting from a simplified scenario, the analysis here is based on the following assumptions:
\begin{itemize}
	\item The condition in \eqref{e:GenCond1} is satisfied;
	\item A packet not in transmission is dropped when a new packet is generated. The scenario where devices have buffers is analyzed in Subsection~\ref{ss:minislotDelayBuffer};
	\item All devices are of the same type and priority. Consequently, the three assignment cycles reduce to a unified frame with $n_\mathrm{s}$ slots;
	\item The SyncCS is not adopted. The analysis of SyncCS is given in Subsection~\ref{ss:SynchSens}.
\end{itemize}

We focus on the delay  analysis since collision probability is zero without SMsA.  Let $\tau_0$ denote the \textit{base delay}, defined as the time duration from the packet arrival instant till the first assigned mini-slot. Under the aforementioned assumptions, the average base delay is equal to $n_\mathrm{s} T_\mathrm{s}/2$ for all devices, as each device has one assigned mini-slot in each frame. The \textit{overall delay} is the base delay plus the \textit{access delay} (AD), i.e., the duration from the first assigned mini-slot since the packet arrival till the end of the packet transmission. Since the average base delay is a constant here, we focus on finding the average AD. 

Denote the device assigned the $m$th mini-slot of the $l$th slot by $d_{m, l}$. Denote by $\tau_{m, l}$ the average \emph{access delay counted in frames} (AD-F), i.e., \textcolor{Black}{the number of logical frames since device $d_{m, l}$'s packet arrival till device $d_{m, l}$'s packet transmission.~\footnote{\textcolor{Black}{When packet re-transmission is considered, the definition of AD-F should be changed to ``the number of logical frames since packet arrival till successful packet transmission''. Meanwhile, the ``packet arrival rate'' in our analysis should be replaced by ``packet transmission rate'' as a packet may need re-transmission(s).}} Different from a physical frame, a logical frame counted in the AD-F for device $d_{m, l}$ is the duration from slot $l$ of one physical frame to slot $l$ of the next physical frame. Therefore, a logical frame has the same length as a physical frame, but different starting and ending points for different devices. Accordingly, the arrival and transmission of a packet can happen within one logical frame, and the resulting AD-F is 1 in such case.}\footnote{ \textcolor{Black}{In the rest of the paper, we do not distinguish physical and logical frames and refer to both as ``frame'' since they are equal in length.}} Note that AD-F $\tau_{m, l}$ corresponds to a duration slightly longer than the AD defined in the preceding paragraph. This is because the AD ends when a packet transmission is completed, while the AD-F counts the entire frame in the delay, including the duration after device $d_{m, l}$'s packet transmission. Accordingly, the AD of device $d_{m, l}$ can be obtained from the AD-F by calculating $(\tau_{m, l} - 1) \times T_\mathrm{f} + T_\mathrm{x}$, where the frame length $T_\mathrm{f}$ is equal to $n_\mathrm{s} T_\mathrm{s}$.

Since any device assigned the first mini-slot of any slot can transmit right away without sensing when the slot begins, we have
\begin{align}
\tau_{1, l} = 1, \forall l.
\end{align}
For devices assigned the subsequent mini-slots, the AD-F can be found using the following result.

\begin{theorem}
	For any integer $m$ such that $1 \leq m \leq n_\mathrm{m}-1$, the following relation between the AD-F of device $d_{m+1, l}$ and device $d_{m, l}$ holds:
	\begin{align}\label{e:delayRelaSolu}
	\tau_{m+1, l} = & \frac{1}{ 1 -  \gamma_{m, l} - T_\mathrm{f}\lambda_{m, l}^\prime } \left( - \frac{(1 - \gamma_{m, l}) T_\mathrm{f} \lambda_{m, l}^\prime }{2}  \tau_{m, l}^2  \right. \nonumber \\ & \left. + \big(1 \!-\! \gamma_{m, l}  \!+\!  T_\mathrm{f} \lambda_{m, l}^\prime \big)\tau_{m, l}  - \frac{T_\mathrm{f}\lambda_{m, l}^\prime  (1 \!+\! \gamma_{m,l})}{2}  \right)
	\end{align}
	where
	\begin{align}\label{e:EqvArrivalRateReplace}
	\lambda_{m, l}^\prime =  \frac{\lambda_{m, l}}{ \big(1 + T_\mathrm{f} \lambda_{m, l}(\tau_{m, l} - 1/2)\big)}
	\end{align}
	represents the effective packet arrival rate of device $d_{m, l}$ excluding dropped packets due to packet replacement (as there is no buffer), and
	\begin{align}
	\gamma_{m, l} &= T_\mathrm{f} \sum\limits_{r = 1}^{m}  \lambda_{r, l}^\prime  
	\end{align}
	represents the expected overall number of packet arrivals in a frame for devices $d_{1, l}$ till $d_{m, l}$ (excluding replaced packets).
\end{theorem}


Using the fact that $\tau_{1, l}$ is equal to 1 for any $l$, \eqref{e:delayRelaSolu} can be used to obtain the AD-F for devices assigned to all subsequent mini-slots in a slot recursively.

\subsection{Delay Performance with Buffer}\label{ss:minislotDelayBuffer}

Now consider the case when each device has a buffer. Recall that different mini-slots correspond to different transmission priorities. In the proposed protocol, any proper slot and mini-slot assignment ensures that the expected number of packets in the buffer of device $d_{m, l}$ is less than one, for any $m < n_\mathrm{m}$ and any $l$. The reason is that, if the expected number of buffered packet at $d_{m, l}$ is larger than or equal to one, devices assigned mini-slots $m+1, \dots, n_\mathrm{m}$ of slot $l$ have almost no opportunity to transmit.
As a result, we neglect the case when there are more than one packet in a buffer and use the following approximation. Specifically, at any instant, a device is in one of three states:
\begin{itemize}
	\item no packet;
	\item one packet, transmitting or waiting for channel access;
	\item two packets, one transmitting or waiting for channel access and the other arriving and going into the buffer.
\end{itemize}
Accordingly, for any given device, there is either no packet or one packet transmitting or waiting for channel access when a new packet arrives.

Denote by $\tau_{m, l}^\mathrm{b}$ the average AD-F of device $d_{m, l}$ in the case with buffer, the following result is in order.

\begin{theorem}
	In the case with buffers, for any integer $m$ such that $1 \leq m \leq n_\mathrm{m}-1$, the relation between the AD-F of device $d_{m+1, l}$ and device $d_{m, l}$ is given by
\begin{align}\label{e:delayRelaBuffer}
\tau^\mathrm{b}_{m+1,l} \!
=& \frac{1 \!-\! \gamma_{m, l}}{1 \!-\! \gamma_{m+1,l} } \!\left(\!\! \frac{1}{ 1 \!-\!  \gamma_{m, l}^\mathrm{b} \!-\! T_\mathrm{f}\lambda_{m, l} } \!\! \left( \!\!- \frac{(1 \!-\! \gamma_{m, l}^\mathrm{b})T_\mathrm{f} \lambda_{m, l}}{2}   \tau_{m, l}^2 \right. \right. \nonumber\\ %
& \left. \left. + (1 \!-\! \gamma_{m, l}^\mathrm{b}  \!+\!  T_\mathrm{f} \lambda_{m, l} )\tau_{m, l}  \!-\! \frac{T_\mathrm{f}\lambda_{m, l} (1 \!+\! \gamma_{m,l}^\mathrm{b})}{2} \! \right)  -\! 1 \!\right) \nonumber \\
& \left. \left.+ 1 \right. \right.
\end{align}
where
\begin{align}
\gamma^\mathrm{b}_{m, l} = T_\mathrm{f}  \sum\limits_{r = 1}^{m}  \lambda_{r, l}
\end{align}
represents the expected overall number of packet arrivals in a frame for devices $d_{1, l}$ till $d_{m, l}$.
\end{theorem}

\subsection{Slot Idle Probability}

A slot is idle if none of its associated devices transmits. Under stationary packet arrival statistics, the expected slot idle probability of MsCS can be obtained. In the case with and without buffer, the slot idle probability is approximately given by
\textcolor{Black}{
\begin{subequations}
\begin{align}
\eta_{l}^\mathrm{b} = 1 - \sum\limits_{m=1}^{n_\mathrm{m}} \lambda_{m,l} T_\mathrm{f} \\
\eta_{l} = 1 - \sum\limits_{m=1}^{n_\mathrm{m}} \lambda_{m,l}^\prime T_\mathrm{f}
\end{align}
\end{subequations}
} 
respectively, where $\lambda^\prime_{m, l}$ is given in \eqref{e:EqvArrivalRateReplace}. \textcolor{Black}{Note that the right-hand side of either of the two equations above is non-negative when the condition \eqref{e:GenCond1} is satisfied, i.e., when the slot is not overloaded. The above approximation of slot idle probability also assumes a negligible packet collision probability, which means the expected number of transmitted packets and the expected number of packet arrivals (that cause no packet replacement) are equal in any slot.}

Define the throughput of slot $l$ as the expected number of packets transmitted in the slot $l$. The slot throughput equals $1 - \eta_{l}^\mathrm{b}$  and $1 - \eta_{l}$ for the cases with and without buffers.

\subsection{Impact of SyncCS}\label{ss:SynchSens}

As SyncCS results in two possible lengths for each slot, i.e., the full and the reduced lengths, the frame length becomes a random variable. Denote the expected frame length with SyncCS in the case with and without buffer by $T^\mathrm{e, b}_\mathrm{f}$ and $T^\mathrm{e}_\mathrm{f}$, respectively.
Denote $n^\prime_\mathrm{s}$ as the number of busy slots out of the $n_\mathrm{s}$ slots in a frame. In the case without buffer, it follows that
\begin{align}\label{e:FrameLen}
T^\mathrm{e}_\mathrm{f} = n_\mathrm{s} n_\mathrm{m}T_\mathrm{m} + n^\prime_\mathrm{s} T_\mathrm{x}.
\end{align}
Since there is no collision,  
\begin{align}\label{e:FrameEqApprox}
T^\mathrm{e}_\mathrm{f}  \sum\limits_l \sum\limits_{m} \lambda_{m, l}^\prime = n^\prime_\mathrm{s}
\end{align}
because the expected number of packet transmissions should equal the expected number of arriving packets (that are not replaced) in the a frame duration. From \eqref{e:FrameLen}, \eqref{e:FrameEqApprox}, and \eqref{e:EqvArrivalRateReplace} (with $T_\mathrm{f}$ replaced by $T^\mathrm{e}_\mathrm{f}$), $n^\prime_\mathrm{s}$ and $T^\mathrm{e}_\mathrm{f}$ can be solved.


In the case with buffers, we have
\begin{subequations}
\begin{align}\label{e:FrameEqApproxBuffer}
T^\mathrm{e, b}_\mathrm{f} = n_\mathrm{s} n_\mathrm{m}T_\mathrm{m} + n^\prime_\mathrm{s} T_\mathrm{x}\\
T^\mathrm{e, b}_\mathrm{f} \sum\limits_l \sum\limits_{m} \lambda_{m, l}   = n^\prime_\mathrm{s}
\end{align}
\end{subequations}
which gives
\begin{align}\label{e:FrameEqApproxBufferEqv}
T^\mathrm{e, b}_\mathrm{f} = \frac{ n_\mathrm{s} n_\mathrm{m}T_\mathrm{m}}{1 - \sum\limits_l \sum\limits_{m} \lambda_{m,l} T_\mathrm{x}}.
\end{align}

Substituting $T_\mathrm{f}$ in \eqref{e:delayRelaSolu} and \eqref{e:delayRelaBuffer} with $T^\mathrm{e}_\mathrm{f}$ and $T^\mathrm{e, b}_\mathrm{f}$, respectively,  gives the AD-F of the proposed design with MsCS and SyncCS. In the case without buffer, $T^\mathrm{e}_\mathrm{f}$ depends on $\tau_{m, l}$ through \eqref{e:EqvArrivalRateReplace}, which renders a complicated relation.

\subsection{Impact of SMsA}\label{ss:SMsA}

The AD-F in Subsections~\ref{ss:minislotDelay}~and~\ref{ss:minislotDelayBuffer} is obtained when each mini-slot is assigned to a device exclusively. With SMsA, we have the following questions:
\begin{itemize}
	\item What is the relation among the AD-F of different devices assigned the same mini-slot?
	\item How does the SMsA impact the relation in the AD-F between devices assigned adjacent mini-slots?
\end{itemize}

Denote the set of all devices assigned mini-slot $m$ of slot $l$ by $\mathcal{D}_{m,l}$. The following theorem answers the first question.
\begin{theorem}\label{T:Th3}
	In the case without buffer, all devices in $\mathcal{D}_{m, l}$ have the same AD-F, \textcolor{Black}{regardless of the difference in their individual packet arrival rates}. In the case with buffer, assuming negligible packet collision probability and
	\begin{align}\label{e:CondTheorem3}
	\lambda_i \ll \sum\limits_{r = 1}^{m} \sum\limits_{j \in \mathcal{D}_{r,l}} \lambda_{j}, \forall i \in \mathcal{D}_{m,l}
	\end{align}
	the differences among the AD-Fs of devices in $\mathcal{D}_{m, l}$ are negligible.
\end{theorem}



For the second question, similar to \eqref{e:delayRelaSolu}~and~\eqref{e:delayRelaBuffer}, the relation between the AD-Fs of devices in adjacent mini-slots in the case of SMsA can be characterized. For brevity, the characterization is given in the proof of Theorem~3 in Section~\ref{AppendixSuperpose}. 
\textcolor{Black}{It is worth mentioning that the packet collision probability has an impact on the AD-F even if devices do not detect collisions or re-transmit. Given the aggregated packet arrival rate of devices sharing a mini-slot, a higher collision probability implies less channel busy duration for transmitting the same amount of packets. Consequently, the average packet waiting time and the AD-F reduce as the collision probability increases. However, if the collision probability is low, such impact can be negligible.}

With the AD-F, we can estimate the packet collision probability. Consider the case with buffer as an example and assume that the condition in \eqref{e:CondTheorem3} is satisfied. Based on Theorem~\ref{T:Th3}, all the devices in $\mathcal{D}_{m, l}$ have the same AD-F, denoted  by $\tau_{m, l}^\mathrm{s, b}$. Then, any device in $\mathcal{D}_{m, l}$ is expected to have one transmission opportunity in every $\tau_{m, l}^\mathrm{s, b}$ frames. The expected number of packet arrivals at device $i\in \mathcal{D}_{m, l}$ between any two consecutive transmission opportunities, which must be less than 1, can be estimated by $\tau_{m, l}^\mathrm{s, b} T_\mathrm{f} \lambda_i$. With the MsCS, all devices in $\mathcal{D}_{m, l}$ that have packets to send share the same transmission opportunities. Therefore, the probability that device $i$'s packet encounters a collision is approximately given by
\begin{align}\label{e:collisonProbEstimateBuffer}
q^\mathrm{c,b}_i = 1 - \prod\limits_{j\in \mathcal{D}_{m, l}\backslash \{i\}} \big(1 - \tau_{m, l}^\mathrm{s,b} T_\mathrm{f} \lambda_i \big).
\end{align}
Note that, knowing only the average packet arrival rates,  the above approximation may be limited in accuracy. An accurate determination of the collision probability requires the traffic arrival model of all devices, which can be difficult to obtain in practice. We demonstrate through numerical results that our approximation can be a useful tool for device assignment in the case with SMsA in Part~II.

\section{Conclusion}\label{s:Conclude}

In Part~I of this paper, we tailor a MAC protocol for MTC in IIoT. To increase channel utilization efficiency, we propose MsCS and SyncCS, both of which feature distributed coordination. To prioritize devices and guarantee the QoS requirement of HP devices, we adopt differentiated assignment cycles for different types of devices. To further increase the supported number of devices, we develop the idea of SMsA, which can multiply the network capacity with a delay-collision trade-off. Thanks to the strategies, the overall protocol has the potential to simultaneously achieve the targets of improving channel usage, minimizing messaging overhead, satisfying stringent QoS constraints, and providing differentiated performance. Meanwhile, customized and effective packet transmission scheduling that complements the protocol is necessary for achieving the full potential of the proposed protocol and is studied in Part~II, which also presents numerical results to evaluate performance of the proposed MAC protocol.   

\appendices\label{sec:Appendix}

\section{Proof of Theorem~1}\label{sec:AppendixDelay}

We first prove the effective packet arrival rate in \eqref{e:EqvArrivalRateReplace}. A packet is subject to replacement in the duration from its arrival till the beginning of its transmission.  Given the average AD-F $\tau_{m, l}$, the average of aforementioned duration is $\tau_{m, l} - 1 + 1/2$ frames, where $-1$ excludes the frame of transmission while $1/2$ adds half frame due to the average base delay. Given packet arrival rate  $\lambda_{m, l}^\prime$ and neglecting the correlation between packet arrivals, the probability that a new packet arrives in this duration can be estimated by
$(\tau_{m, l} - 1/2)/ (1/(n_\mathrm{s} T_\mathrm{s} \lambda^\prime_{m, l}))$, where $1/(n_\mathrm{s} T_\mathrm{s} \lambda^\prime_{m, l})$ is the average number of frames per packet arrival. Therefore, the following equation regarding packet arrival and replacement holds:
\begin{align}
\lambda_{m, l}^\prime = \lambda_{m, l} (1 - n_\mathrm{s} T_\mathrm{s} \lambda^\prime_{m, l}(\tau_{m, l} - 1/2))
\end{align}
which gives \eqref{e:EqvArrivalRateReplace}.

Suppose that device $d_{m, l}$ has a packet ready to transmit at the beginning of slot $l$. We have the following observations:
\begin{itemize}
	\item  Device $d_{m+1, l}$ has the same AD-F as device $d_{m, l}$ if: i) device $d_{m, l}$ is removed from mini-slot $m$, and ii) the packet arrival probability of  $d_{m+1, l}$ is the same as the packet arrival probability of device  $d_{m, l}$;
	\item Consider independent packet arrivals at devices $d_{m, l}$ and $d_{m+1, l}$. Even if the packet arrival rates of devices $d_{m, l}$ and $d_{m+1, l}$ are different, the following two probabilities are the same,  given any realizations of the packet arrival processes of devices $d_{1, l}$ to $d_{m - 1, l}$ and for any arbitrary~$x\geq 0$. The first is the probability that the AD-F of a packet of device $d_{m, l}$ is $x$, while the second is the probability that the AD-F of a packet of device $d_{m+1, l}$ is $x$ if device $d_{m, l}$ is removed from the slot;
	\item If a packet of device $d_{m+1, l}$ arrives when a packet of device $d_{m, l}$ is waiting for channel access, then the packet of device $d_{m+1, l}$ needs to wait till at least one frame after the packet of device $d_{m, l}$ is sent. During the waiting in such case, it is possible that there are new packet arrivals at devices $d_{0, l}, \dots, d_{m, l}$, which can trigger further waiting for the packet of device  $d_{m+1, l}$.
\end{itemize}

Accordingly, for devices assigned to mini-slot $m$ ($1 \leq m \leq n_\mathrm{m}-1$), we can obtain the following relation between $\tau_{m+1, l}$ and $\tau_{m, l}$:
\begin{align}\label{e:delayRela}
\tau_{m+1, l} =&  \bigg( 1 - \alpha_{m, l} - n_\mathrm{s} T_\mathrm{s} \lambda_{m, l}^\prime - \beta_{m, l} \bigg) \tau_{m, l} \nonumber \\
& +  \alpha_{m, l} \left(\left(\frac{\tau_{m, l} - 1}{2} + 1\right)\frac{1}{1 - \gamma_{m, l}} + 1 \right) \nonumber \\
& + n_\mathrm{s} T_\mathrm{s} \lambda_{m, l}^\prime \left( 1 + \frac{1}{1 - \gamma_{m, l}}\right) \nonumber \\
& + \beta_{m, l} \bigg(\tau_{m, l} + \frac{1}{1 - \gamma_{m, l}}\bigg)
\end{align}
where
\begin{align}\label{e:alpha_ml}
\alpha_{m, l} = \frac{  \tau_{m, l} - 1} {1/(n_\mathrm{s} T_\mathrm{s} \lambda_{m, l}^\prime)} = (\tau_{m, l} - 1) n_\mathrm{s} T_\mathrm{s} \lambda_{m, l}^\prime
\end{align}
represents the probability that a packet of device $d_{m+1, l}$ arrives while device $d_{m, l}$ has a packet waiting for channel access but not transmitting. The probability is approximated by the ratio of the average number of frames for device $d_{m+1, l}$'s packet to wait to the average number of frames between two packet arrivals of $d_{m+1, l}$. The parameter $\beta_{m, l}$ in \eqref{e:delayRela} is given by
\begin{align}\label{e:beta_ml}
\beta_{m, l} =  \frac{\tau_{m+1, l} - 1} {1/(n_\mathrm{s} T_\mathrm{s} \lambda_{m, l}^\prime)},
\end{align}
and it represents the probability that a packet of device $d_{m+1, l}$ arrives when device $d_{m, l}$ has no packet yet, but device $d_{m, l}$ will have a packet arrival and transmit that packet before device $d_{m+1, l}$ transmits its packet.

The four terms on the right-hand side of \eqref{e:delayRela} correspond to the following four cases:
\begin{itemize}
	\item With probability $1 - \alpha_{m,l} - n_\mathrm{s} T_\mathrm{s} \lambda_{m, l} - \beta_{m, l}$, device $d_{m, l}$ has no packet waiting for channel access between the arrival and transmission of a packet of device $d_{m+1, l}$. In such case, the expected AD-F of this packet of device $d_{m+1, l}$ is the same as the expected AD-F of a packet of device $d_{m, l}$;
	\item With probability $\alpha_{m, l}$, a packet of device $d_{m+1, l}$ arrives when a packet of device $d_{m, l}$ is waiting for channel access (but not transmitting). In such case, the  AD-F for the packet of $d_{m+1, l}$ is equal to the number of frames device $d_{m, l}$ needs to wait for from now on, added by one more frame plus the average number of packet arrivals at the devices from $d_{0, l}$ till $d_{m, l}$;
	\item With probability $n_\mathrm{s} T_\mathrm{s} \lambda_{m, l}^\prime$, a packet of device $d_{m+1, l}$ arrives when device $d_{m, l}$ is transmitting. In such case, there was no packet waiting for channel access at devices assigned to precedent mini-slots when the packet transmission started. Therefore, the expected AD-F is equal to the expected number of packet arrivals of devices $d_{0, l}$ to $d_{m, l}$ in this frame plus expected number of packets arrivals during the transmission of these packets plus one frame of transmission time;
	\item With probability $\beta_{m, l}$, a packet of device $d_{m+1, l}$ arrives when $d_{m, l}$ has no packet waiting or transmitting, and then a packet of device $d_{m, l}$ arrives while the packet of device $d_{m+1, l}$ is waiting for channel access. \textcolor{Black}{In such case, compared to the expected AD-F of a new packet at device $d_{m+1, l}$ with no packet at $d_{m, l}$, which is equal to $\tau_{m, l}$ based on the argument in the first case above, the waiting time of $d_{m+1, l}$ is increased by 1 (for device $d_{m, l}$} to transmit its packet) plus the average number of packet arrivals at devices $d_{0, l}$ to $d_{m, l}$.
\end{itemize}

Substituting \eqref{e:alpha_ml} and \eqref{e:beta_ml} into \eqref{e:delayRela}, $\tau_{m, l}$ can be found as in \eqref{e:delayRelaSolu}. \hfill $\blacksquare$

\section{Proof of Theorem~2}\label{sec:AppendixDelayBuffer}

Consider the conditional probability that one packet is waiting for channel access given that a new packet arrives at device $d_{m, l}$ in the scenario with buffers. When the packet arrival is independent on the packet transmission, the above probability is the same as the probability that one packet is transmitting or waiting for channel access at device $d_{m, l}$. Denote this probability by $p_{m, l}^\mathrm{b}$. Then, for mini-slot 1, the AD-F, i.e., $\tau_{1,l}^\mathrm{b}$, is 1 with probability $1 - p_{1, l}^\mathrm{b}$. With probability $p_{1, l}^\mathrm{b}$, a new packet of $d_{1, l}$ arrives in the frame in which the existing packet of $d_{1, l}$ is or will be transmitting. The average AD-F $\tau_{1,l}^\mathrm{b}$ in the latter case is 1.5 frames, with 1 transmission frame and an average of 0.5 waiting frame.

The overall average AD-F is given by
\begin{align}
\tau^\mathrm{b}_{1,l} = (1 - p_{1, l}^\mathrm{b})  + 1.5p_{1, l}^\mathrm{b} = 1 + 0.5p_{1, l}^\mathrm{b}.
\end{align}
Under the aforementioned approximation of at most 1 packet in buffer, the probability $p_{m, l}^\mathrm{b}$ can be estimated as 
\begin{align}
p_{1, l}^\mathrm{b} = (\tau^\mathrm{b}_{1,l} - 0.5) \lambda_{1,l} T_\mathrm{f}
\end{align}
where $-0.5$ corresponds to deducting the frame of transmission and adding the 0.5 frame of base delay.
From the above two equations, the average AD-F can be derived as  
\begin{align}
\tau^\mathrm{b}_{1,l} = 1+ \frac{\lambda_{1,l} T_\mathrm{f}}{2(2 - \lambda_{1,l} T_\mathrm{f})}.
\end{align}
For subsequent mini-slots, the average AD-F when a packet arrives with no existing packet at the same device, denoted by $\hat{\tau}^\mathrm{b}_{m,l}$, can be found using the same approach as in the case with no buffer. Meanwhile, with probability $p_{m, l}^\mathrm{b}$, a packet arrives when another packet is waiting for channel access, which  will experience the  average AD-F given by
\begin{align}
\tilde{\tau}^\mathrm{b}_{m,l} = \hat{\tau}^\mathrm{b}_{m,l}  +  \frac{1}{1 - \gamma^\mathrm{b}_{m,l}}.
\end{align}
Thus, the overall average AD-F is 
\begin{align}\label{e:tauml_b}
\tau^\mathrm{b}_{m,l} &= p_{m, l}^\mathrm{b} (\hat{\tau}^\mathrm{b}_{m,l} + \frac{1}{1 - \gamma_{m,l}}) + (1 - p_{m, l}^\mathrm{b}) \hat{\tau}^\mathrm{b}_{m,l} \nonumber \\
& = \hat{\tau}^\mathrm{b}_{m,l} + p_{m, l}^\mathrm{b}  \frac{1}{1 - \gamma_{m,l}}.
\end{align}
The probability, $p_{m, l}^\mathrm{b}$, can be found as
\begin{align}\label{e:pml_b}
p_{m, l}^\mathrm{b} = \frac{\tau^\mathrm{b}_{m,l}-1} {1/(\lambda_{m,l} T_\mathrm{f})}.
\end{align}
Similarly, from \eqref{e:tauml_b}~and~\eqref{e:pml_b}, it can be found that
\begin{align}\label{e:DelayBufferRela}
\tau^\mathrm{b}_{m,l} &= \frac{\hat{\tau}^\mathrm{b}_{m,l} - 1}{1 - \lambda_{m,l} T_\mathrm{f}  \frac{1}{1 - \gamma_{m,l}}} + 1  \nonumber \\
&= \frac{1 - \gamma_{m, l}}{1 - \gamma_{m+1,l} } (\hat{\tau}^\mathrm{b}_{m,l} - 1) + 1.
\end{align}
Using \eqref{e:delayRelaSolu}, but with the effective packet arrival rate $\lambda^\prime_{m,l}$ replaced by $\lambda_{m,l}$ (since there is no packet replacement in the case with buffer), it follows that
\begin{align}\label{e:delayRelaBufferInt}
\hat{\tau}^\mathrm{b}_{m+1, l}
=& \frac{1}{ 1 \!-\!  \gamma_{m, l}^\mathrm{b} \!-\! T_\mathrm{f}\lambda_{m, l} } \left( - \frac{1}{2} \left(1 - \gamma_{m, l}^\mathrm{b} \right) T_\mathrm{f} \lambda_{m, l} \tau_{m, l}^2 \right. \nonumber \\
& \left. + (1 \!-\! \gamma_{m, l}^\mathrm{b}  \!+\!  T_\mathrm{f} \lambda_{m, l} )\tau_{m, l}  \!-\! \frac{1}{2} T_\mathrm{f}\lambda_{m, l} (1 \!+\! \gamma_{m,l}^\mathrm{b})  \right).
\end{align}

Using \eqref{e:DelayBufferRela} and \eqref{e:delayRelaBufferInt}, the average AD-F for mini-slot $m \geq 2$ in \eqref{e:delayRelaBuffer} can be obtained. \hfill $\blacksquare$

\section{Proof of Theorem~3}\label{AppendixSuperpose}

\subsubsection{The Case with No Buffer}

Denote by $q^\mathrm{c}_i$ the conditional packet collision probability of device $i$ given that device $i$ is transmitting. Note that, as each mini-slot is assigned to multiple devices, we cannot use $d_{m,l}$ to represent the device assigned to mini-slot $m$ of slot $l$. Because of a non-zero collision probability, the expected number of transmitting packets given that device $i$ is transmitting, denoted by $n^\mathrm{c}_i$, can be larger than 1.

Let $\Lambda_{m, l}^\prime$ denote the aggregated effective packet arrival rate of all devices assigned to mini-slot $m$ of slot $l$, given by
\begin{align}\label{e:SumRateSup}
\Lambda_{m, l}^\prime &=  \sum\limits_{i \in \mathcal{D}_{m,l}} \lambda_{i}^\prime \bigg((1 - q^\mathrm{c}_i) \cdot 1 + q^\mathrm{c}_i \cdot \frac{n^\mathrm{c}_i - 1}{n^\mathrm{c}_i}\bigg)   \nonumber \\
&= \sum\limits_{i \in \mathcal{D}_{m,l}} \lambda_{i}^\prime \bigg(1 - \frac{q^\mathrm{c}_i}{n^\mathrm{c}_i}\bigg)
\end{align}
where $\lambda_{i}^\prime$ is the effective arrival rate in \eqref{e:EqvArrivalRateReplace}. The impact of packet replacement and packet collision is considered in \eqref{e:SumRateSup}, where all packets simultaneously transmitting in the same mini-slot are only counted as one. While $q^\mathrm{c}_i$ and $n^\mathrm{c}_i$ are conditional (on a packet transmission of device $i$), $\lambda_{i}^\prime$ implicitly indicates the packet transmission rate of device $i$ (as each packet taken into account by the effective arrival rate will be transmitted). As a result, $\Lambda_{m, l}^\prime$ is unconditional.

Denote the average cumulative effective packet arrival rate for all devices assigned to the first $m$ mini-slots by
\begin{align}\label{e:GammaSup}
\Gamma_{m, l} = T_\mathrm{f}\sum\limits_{r = 1}^{m} \Lambda_{m, l}^\prime.
\end{align}
Given $\Lambda_{m, l}^\prime, \forall m, l$, the average AD-F in the case of SMsA can be found by extending the result in \eqref{e:delayRelaSolu} as follows:
\begin{align}\label{e:delayRelaSup}
\tau_{m+1, l}^\mathrm{s} \!=& \frac{1}{ 1 -  \Gamma_{m, l} - T_\mathrm{f}\Lambda_{m, l}^\prime } \left( - \frac{ \left(1 - \Gamma_{m, l} \right) T_\mathrm{f} \Lambda_{m, l}^\prime}{2} (\tau_{m, l}^\mathrm{s})^2 \right. \nonumber \\
& + \! \left. (1 \!-\! \Gamma_{m, l} \!+\!  T_\mathrm{f} \Lambda_{m, l}^\prime )\tau_{m, l}^\mathrm{s}  - \frac{1}{2} T_\mathrm{f}\Lambda_{m, l}^\prime  (1 + \Gamma_{m,l})  \right).
\end{align}

With $\tau_{m, l}^\mathrm{s}$, the packet collision probability for device $i$ in $\mathcal{D}_{m,l}$ can be approximated by
\begin{align}\label{e:collisonProbEstimate}
q^\mathrm{c}_i = 1 - \prod\limits_{j\in \mathcal{D}_{m, l}\backslash \{i\}} \big(1 - \tau_{m, l}^\mathrm{s} T_\mathrm{f} \lambda_i \big).
\end{align}
Meanwhile, $n^\mathrm{c}_i$ can be estimated by
\begin{align}\label{e:collisonNumEstimate}
n^\mathrm{c}_i = 1 +  \sum\limits_{j\in \mathcal{D}_{m, l}\backslash \{i\}} \tau_{m, l}^\mathrm{s} T_\mathrm{f} \lambda_j,
\end{align}
where constant 1 corresponds to the given fact that a packet of device $i$ is involved in the collision.

Using \eqref{e:SumRateSup}~to~\eqref{e:collisonNumEstimate} and the fact that $\tau_{1, l}^\mathrm{s} = 1, \forall l$, the average AD-F and packet collision probability for all devices assigned to all mini-slots can be derived. The result in \eqref{e:delayRelaSup} suggests that, although the packet arrival rates for devices assigned to one mini-slot can be different, the average AD-Fs for the devices are the same. This is not unexpected if we compare \eqref{e:delayRelaSolu}~and~\eqref{e:delayRelaSup}. From the comparison, it can be seen that the impact of individual packet arrival rate on the average AD-F is replaced by the impact of aggregated packet arrival rate of the mini-slot in the case of SMsA. \textcolor{Black}{Thus, while individual devices may have different packet arrival rates, their average AD-Fs become identical as all devices assigned to the same mini-slot share the same aggregated packet arrival rate}.


\subsubsection{The Case with Buffers}

Denote by $q^\mathrm{c, b}_i$ the conditional packet collision probability of device $i$ given that device $i$ is transmitting. Because of a non-zero collision probability, the expected number of transmitting packets given that device $i$ is transmitting, denoted by $n^\mathrm{c, b}_i$, can be larger than 1.

Let $\Lambda_{m, l}^\mathrm{b}$ denote the aggregated effective packet arrival rate of all devices assigned to mini-slot $m$ of slot $l$, given by
\begin{align}\label{e:SumRateSupBuffer}
\Lambda_{m, l}^\mathrm{b} = \sum\limits_{i \in \mathcal{D}_{m,l}} \lambda_{i} (1 - \frac{q^\mathrm{c, b}_i}{n^\mathrm{c, b}_i}).
\end{align}
Denote the cumulative effective packet arrival rate for all devices assigned to the first $m$ mini-slots by
\begin{align}\label{e:GammaSupBuffer}
\Gamma_{m, l}^\mathrm{b} = T_\mathrm{f} \sum\limits_{r = 1}^{m} \Lambda_{m, l}^\mathrm{b}.
\end{align}
Following the analysis leading to the results in \eqref{e:delayRelaSolu}, \eqref{e:delayRelaBuffer} and \eqref{e:delayRelaSup}, the average AD-F for device $i$ assigned to the $m$th mini-slot of slot $l$ can be obtained:
\begin{align}\label{e:delayRelaSupBuffer}
\tau_{i, m+1, l}^\mathrm{s, b}
=& \frac{1 \!-\! \Gamma_{m, l}^\mathrm{b}}{1 \!-\! \Gamma_{m,l}^\mathrm{b} \!-\! T_\mathrm{f} \lambda_{i} } \left(\frac{1}{ 1 \!-\!  \Gamma_{m, l}^\mathrm{b} \!-\! T_\mathrm{f}\Lambda^\mathrm{b}_{m, l} } \left( \! -  \frac{1}{2} \left(1  \!-\! \Gamma_{m, l}^\mathrm{b} \right)  \right.\right. \nonumber  \\
& \left.\left.  \cdot \; T_\mathrm{f} \Lambda_{m, l}^\mathrm{b} (\bar{\tau}_{m, l}^\mathrm{s,b})^2  + \big(1- \Gamma_{m, l}^\mathrm{b}  +  T_\mathrm{f} \Lambda_{m, l}^\mathrm{b} \big)\bar{\tau}_{m, l}^\mathrm{s,b} \right. \right. \nonumber\\
&  - \left.  \frac{T_\mathrm{f}\Lambda_{m, l}^\mathrm{b} (1 + \Gamma_{m,l}^\mathrm{b})}{2}  \bigg)  - 1 \right) + 1
\end{align}
where
\begin{align}
\bar{\tau}_{m, l}^\mathrm{s,b} = \frac{1}{|\mathcal{D}_{m,l}|} \sum\limits_{i \in \mathcal{D}_{m,l}} \tau_{m, l}^\mathrm{s,b}
\end{align}
represents the average AD-F of all devices assigned into the $m$th mini-slot of slot $l$. From \eqref{e:delayRelaSupBuffer} it can be seen that, different from the case in \eqref{e:delayRelaSup} where the average AD-F is identical for all devices assigned to the same mini-slot, the average AD-F can be different for different devices here due to  $-T_\mathrm{f}\lambda_i$ in the denominator of the first term. However, when the collision probability is low and the condition in \eqref{e:CondTheorem3} is satisfied, the difference by $T_\mathrm{f}\lambda_i$ is negligible in comparison with $\Gamma_{m,l}^\mathrm{b}$. 

With $\tau_{m, l}^\mathrm{s, b}$, $q^\mathrm{c, b}_i$ and $n^\mathrm{c, b}_i$ can be estimated similarly as in the case without buffer.  \hfill $\blacksquare$

%
%

\ifCLASSOPTIONcaptionsoff
  \newpage
\fi


%

\medskip

\balance

\bibliographystyle{IEEEtran}

%
%
%
%
%
%
%

\end{document}